\documentclass[%
 reprint,
superscriptaddress,
 amsmath,amssymb,
 aps,
floatfix,
]{revtex4-2}

\usepackage{graphicx}
\usepackage{dcolumn}
\usepackage{bm}
\usepackage{lipsum}
\usepackage{float}
\usepackage{hyperref}
\hypersetup{
   colorlinks=true,
   linkcolor=blue,
   filecolor=magenta,      
   urlcolor=cyan,
   pdftitle={Overleaf Example},
   pdfpagemode=FullScreen,
   }

\begin{document}

\preprint{APS/123-QED}

\title{Master-slave synchronization of silicon optomechanical nanobeam oscillators by external feedback}

\author{David Alonso-Tomás}
\affiliation{MIND-IN2UB, Departament d'Enginyeria Electrónica i Biomédica, Facultat de Física, Universitat de Barcelona, Martí i Franquès 1, Barcelona 08028, Spain}%
\author{Néstor E. Capuj}
\affiliation{Depto. F\'{i}sica, Universidad de La Laguna, 38200 San Crist\'{o}bal de La Laguna, Spain}%
\affiliation{Instituto Universitario de Materiales y Nanotecnolog\'{i}a, Universidad de La Laguna, 38071 Santa Cruz de Tenerife, Spain}%
\author{Laura Mercadé}
\affiliation{Nanophotonics Technology Center, Universitat Politècnica de València, Camino de Vera s/n, 46022 Valencia, Spain}%
\author{Amadeu Griol}
\affiliation{Nanophotonics Technology Center, Universitat Politècnica de València, Camino de Vera s/n, 46022 Valencia, Spain}%
\author{Alejandro Martínez}
\affiliation{Nanophotonics Technology Center, Universitat Politècnica de València, Camino de Vera s/n, 46022 Valencia, Spain}%
\author{Daniel Navarro-Urrios}
\email{dnavarro@ub.edu}
\affiliation{MIND-IN2UB, Departament d'Enginyeria Electrónica i Biomédica, Facultat de Física, Universitat de Barcelona, Martí i Franquès 1, Barcelona 08028, Spain}%
%
%

\date{\today}

\begin{abstract}
The remote synchronization of oscillators is essential for improving the performance, efficiency, and reliability of various systems and technologies, ranging from everyday telecommunications to cutting-edge scientific research and emerging technologies. In this work, we unequivocally demonstrate a master-slave type of synchronization between two self-sustained optomechanical crystal oscillators that interact solely through an external optical feedback stage. Several pieces of experimental evidence rule out the possibility of resonant forcing, and, in contrast to previous works, indicate that synchronization is achieved in the regime of natural dynamics suppression. Our experimental results are in agreement with the predictions of a numerical model describing the specific mechanical lasing dynamics of each oscillator and the unidirectional interaction between them. The outcomes of our study pave the way toward the synchronization of clock signals corresponding to far-placed processing elements in a future synchronous photonic integrated circuit.
\end{abstract}

\maketitle

Synchronization is the term used to describe the coordination of the temporal dynamics of two or more self-sustained oscillators by means of a weak interaction \cite{Balanov}. First proposed by Lord Huygens in 17th century \cite{Huygens}, this phenomena has been widely found through nature from microscopic to macroscopic world \cite{fussenegger, Hammond, Glass, Lamb, Pikovsky}. Unsurprisingly, synchronization, either unidirectional or bidirectional, has garnered significant interest in the last decades being applied in signal-processing \cite{Heinrich}, RF communications \cite{Bregni}, clock synchronization \cite{Rhee} or even neural networks  \cite{Dorfler} among others. With the advances in micro and nanotechnologies, efforts have been dedicated to achieve the synchronization of self-oscillating micro and nanoelectromechanical systems (MEMS/NEMS), which offer robust high frequency operation, miniaturization and a great scalability  \cite{Asadi,Matheny, Holmes, Aspelmeyer, Roukes}. Optomechanical oscillators (OMOs) are a subset of MEMS/NEMS oscillators that activate large amplitude coherent mechanical motion driven by optical forces\cite{Aspelmeyer}. OMOs are great candidates to explore synchronization mechanisms since the interaction can be mediated by optical signals and, therefore, be effective even at large distances between nodes. However, there are only few studies on this topic performed on OMOs based on microdisk or microspheres resonators \cite{Li, Shah, Shah2, Jang}. On the other hand, there are several proposals concerning synchronization of OMOs that go from purely mechanical synchronization through a mechanical link \cite{Colombano} to the use of a common optical mode that drive both oscillators \cite{Zhang, Zhang2, Bagheri}. These alternatives provide low scalability and would not allow synchronization on demand of distant subsystems within a complex OM network.  
\\
In this manuscript, we unequivocally demonstrate the synchronization between two OMOs based on silicon optomechanical (OM) crystal nanobeams driven by optical forces, which are spectrally separated both in the mechanical and optical domain. The synchronization scheme is a unidirectional master-slave configuration where the light modulation generated by one OMO is fed to the other one. Unlike other circular resonators like disk, spheres, or toroids, OM crystal nanobeams offer a compact and easily integrable solution within a silicon-based platform. These nanobeams are physically connected to the rest of the chip, allowing for direct extraction of coherent mechanical signals into a phononic circuit when needed. Furthermore, the route towards synchronization is performed in the regime of suppression of the natural dynamics described by Balanov \cite{Balanov} instead of the phase-locking mechanisms explored in previous literature \cite{Shah, dani}.\\
\\
An essential condition for synchronization between two oscillators is that both of them should be self-sustained \cite{Balanov, Pikovsky, Matheny}. This means that, in each oscillator, gain overcomes mechanical losses and the mechanical motion becomes coherent and of large amplitude. This regime will be referred hereon as mechanical lasing given its similarity with the optical counterpart and the OM crystal nanobeams will be treated as OMOs. The OM crystals under study are driven to the mechanical lasing regime using a self-pulsing (SP) mechanism that has been explored in previous works \cite{Johnson, dani2} in the sideband unresolved regime. 
 \begin{figure*}[t!]
    \centering    
    \includegraphics[width = 0.95\linewidth]{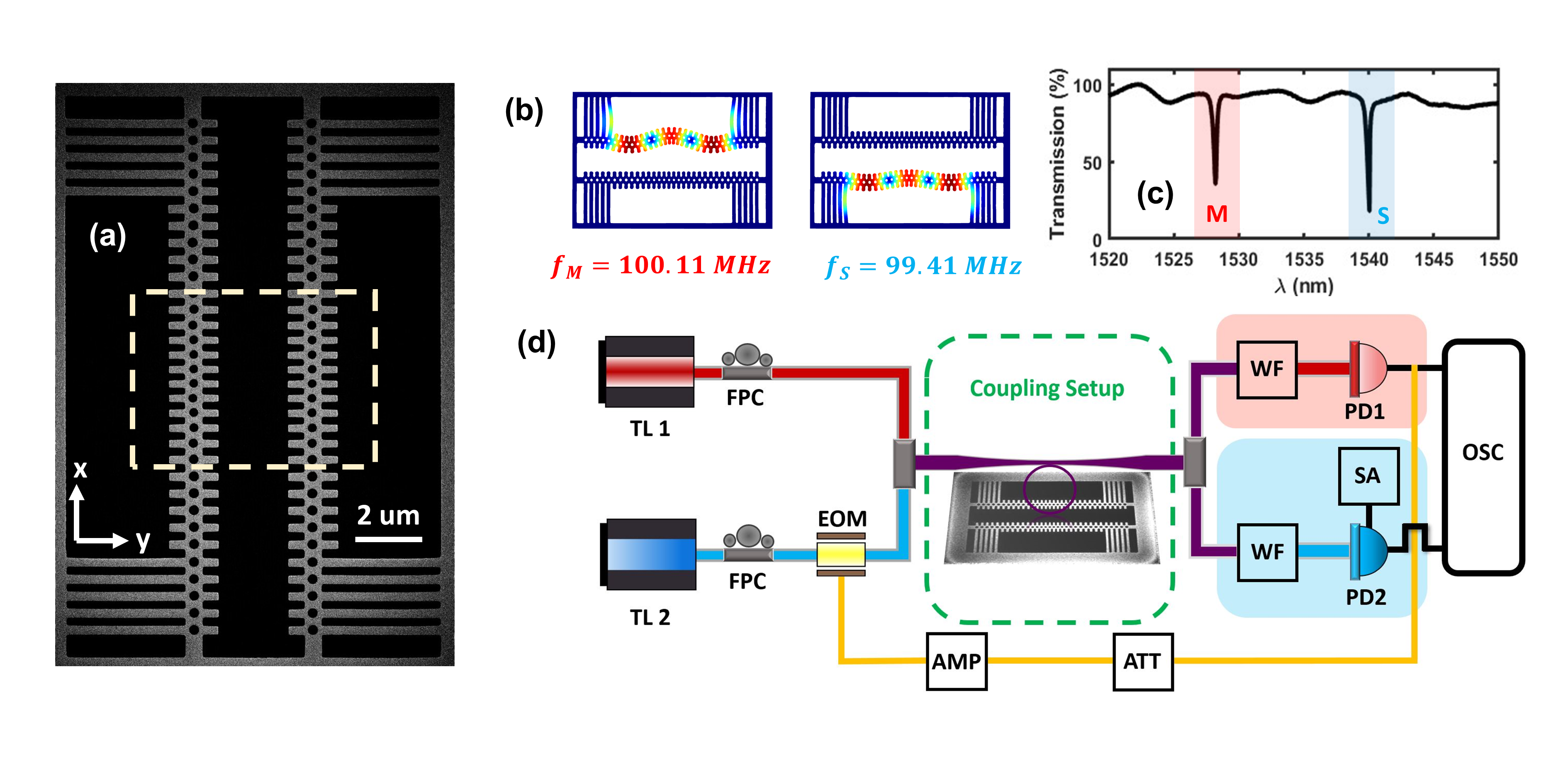}
    \caption{Characteristics of tested optomechanical crystal cavities and experimental setup. \textbf{(a)} SEM image of a pair integrated optomechanical cavities. Optical cavity modes lie in the highlighted region. \textbf{(b)} FEM simulations of the mechanical displacement field of the mechanical mode under study. This analysis utilize a geometry imported from the SEM imageof panel a. \textbf{(c)} Optical transmission spectra of the device under test. The red and blue shaded regions denote master and slave resonances respectively. \textbf{(d)} Schematic of the experimental setup. Tunable laser (TL); fiber polarizer controler (FPC); tunable fabry-perot filter (WF), photodetector (PD); spectrum analyzer (SA); oscilloscope (OSC); attenuator (ATT), amplificator (AMP); electro-optic modulator (EOM). Red and blue paths indicate the different laser wavelengths necessary to excite master and slave resonances, respectively. Purple path indicates the zones where both wavelengths propagate simultaneously.  }
    \label{fig: setup}
\end{figure*}
It is based on the anharmonic modulation of radiation pressure force induced by the dynamical self-limit cycle generated between free-carrier dispersion (FCD) and the thermo-optic (TO) effect in silicon. Essentially, these effects produce a modulation of the refractive index of the material and therefore, a movement of the cavity resonance at a frequency ($\nu_{sp}$). When an harmonic (M) of $\nu_{sp}$ is partially resonant with a mechanical mode of the structure, it can provide coherent amplification and drive it to the mechanical lasing regime. The SP frequency can be thermally tuned in the MHz range by increasing the average intra-cavity photon number ($n_0$) so that the cavity is heated up (see Supplemental Section S2).
The device investigated here is composed of a pair integrated nominally identical one-dimensional OM crystal cavities, which have been fabricated using standard Si nanofabrication techniques on a silicon-on-insulator wafer (see Supplementary Section S1). The outermost five cells on each side of the OM crystals are anchored with tethers to the partially underetched Si frame (Fig. \ref{fig: setup}a). This arrangement ensures that the flexural modes within the plane are isolated from the frame and restricted to the central area of the cavities, which are specifically engineered to sustain high quality factor optical modes for transversal electric (TE) polarization around 1.53 $\mu$m \cite{Gomis}. Both cavities are separated 2 $\mu$m in such a way that it is possible to optically excite them simultaneously by placing a single tapered fiber placed in between. Although these geometries are nominally identical, fabrication imperfections produce slightly different mechanical and optical resonant frequencies. In particular, the mechanical modes used in this work correspond to the in plane flexural ones having 3 anti-nodes along the x direction and mechanical frequencies of ($f_m$, $f_s$) =  ($\Omega_m$, $\Omega_s$)/2$\pi$ = (100.11, 99.41) MHz, where the sub-index m and s denote master and slave, respectively (Fig. \ref{fig: setup}b). The transmission spectrum of the whole device is obtained performing a sweep in wavelength for low input power ($P_{in}$ = 0.5 mW), which results in two well separated optical resonances at ($\lambda_m$, $\lambda_s$) = (1528.20, 1540.05) nm, holding an overall quality factor of $Q_{m,s}$ = (5.68, 6.67) $\cdot$ $10^3$ respectively (Fig. \ref{fig: setup}c). It is worth noting that separated optical resonances are essential to avoid optical cross-talk between the OMOs. \\
\begin{figure*}[t]
    \centering    
    \includegraphics[width = \linewidth]{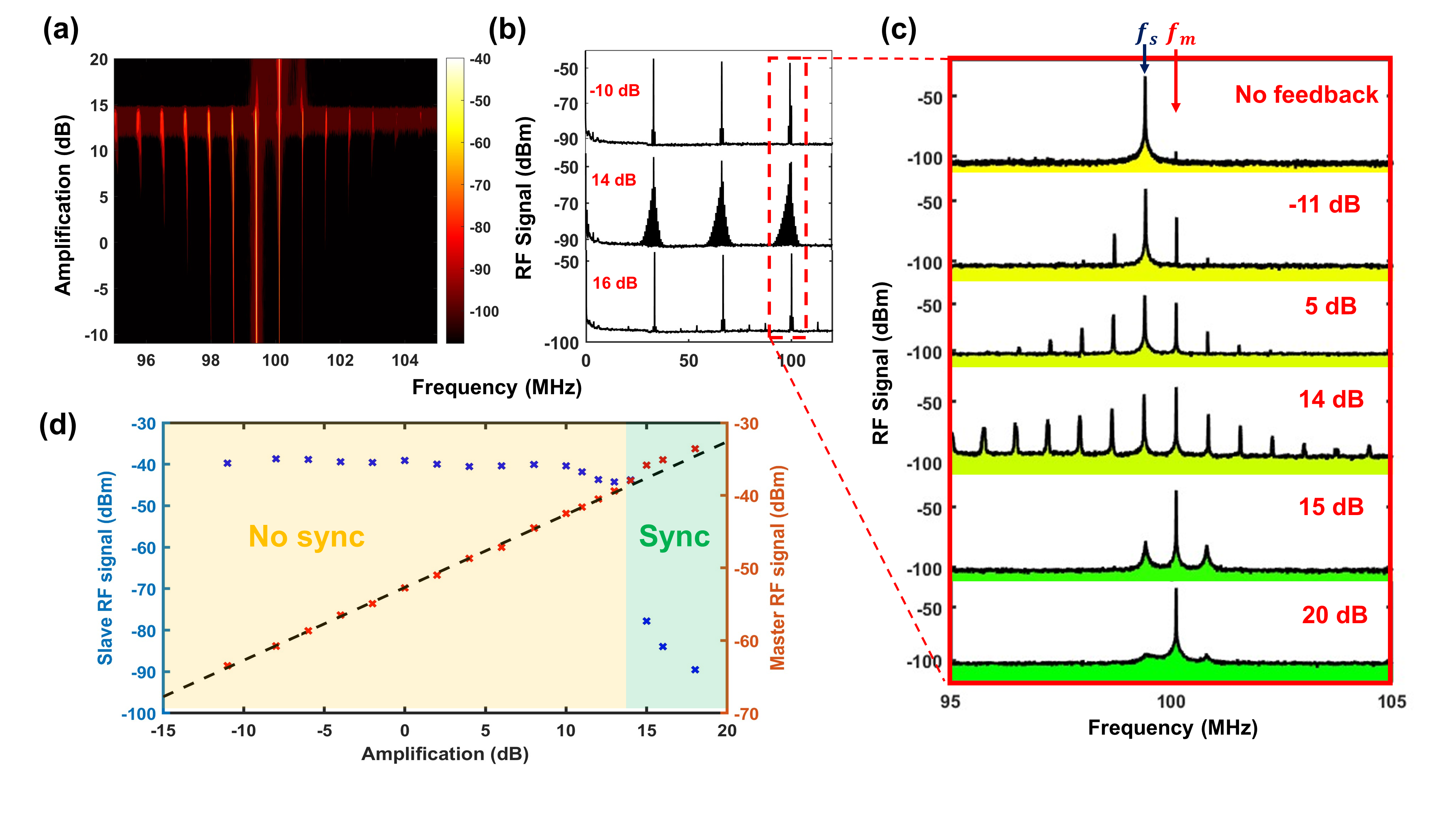}
    \caption{Radio frequency (RF) analysis of the light transmission modulated by the slave dynamics detected at PD2. \textbf{(a)} Contour RF plot near the mechanical natural peak of the slave dynamics as a function of the amplification introduced in the feedback stage. \textbf{(b)} Wider RF spectrum for different amplification power. Note that M = 3 dynamics is present as two extra peaks at one and two thirds of the natural frequency of the slave. \textbf{(c)} Magnification around the natural resonance of the slave. \textbf{(d)} Amplitude of slave's oscillation at slave (blue) and master (red) natural frequency represented as a function of the feedback amplification.}
    \label{fig: res1}
\end{figure*}
Fig. \ref{fig: setup}d shows the setup used to perform the experiment. To achieve the simultaneous optical excitation of both OMOs we employ two tunable lasers, each one tuned at the resonant wavelength of its corresponding OMO. The polarization of each laser is controlled to be TE, which matches that of the cavity modes. Light of the two lasers is afterwards combined and driven to a microloop-shaped fiber that has been thinned down to a diameter of about 1.5 $\mu$m. The bottom part of the microloop can act as a probe that enables the local excitation of the optical cavity modes when the cavities are placed in the near field region of the fiber. Then, light is divided in two paths and spectrally filtered by tunable fabry-perot filters (WF) to record the laser wavelengths resonant with the master and salve cavities in photodetectors PD1 and PD2, respectively. The slave signal is derived to a spectrum analyzer (SA) while the one of the master is introduced as a modulation feedback on the laser exciting the slave by means of an electro-optic modulator (EOM) with a half-wave voltage $V_\pi = 3.5$ V. The offset voltage is set near the quadrature point $V_{DC} = 0.5V_\pi$ so that if the RF modulation signal is small, the output light power responds linearly (see Supplementary Section S3). The magnitude of the feedback signal and, consequently, the modulation amplitude of the slave laser are governed by a stage that allows for tunable attenuation or amplification. As a result of this experimental configuration, the external feedback is the only interaction between both OMOs. Thus, even if the two OMOs are physically placed close to each other, the system is equivalent to have them separated in space. Finally, both detectors output signals are temporally analyzed in a 4-channel oscilloscope (OSC).
\\
By using the SP mechanism explained before, both cavities are excited to different mechanical lasing regimes: M = 3 for the case of the slave (where the third harmonic of the SP is providing the mechanical amplification) and M = 1 for the master. This mechanical lasing scheme allows clearly distinguishing between a forcing mechanism and master-slave synchronization, since the differences may be subtle. 
Indeed, the modulation feedback generated by the master could resonantly drive the slave while eliminating its self-sustained mechanical oscillation. This would obviously lead to a slave mechanical oscillation that will be coherent with that of the master, which may be confused with synchronization. By keeping an M=3 regime in the slave throughout the whole set of measurements we ensure ruling out the possibility of resonant forcing, since otherwise the SP dynamics would disappear.
Fig \ref{fig: res1}a shows the RF response of the slave OMO in a 10 MHz spectral range around the mechanical resonance when varying the feedback amplification. As the feedback modulation amplitude is increased, sidebands resulting from the coherent sum of both non-linear oscillations (slave natural frequency and master modulation harmonics) appear at various possible combination of their frequencies. 
\begin{figure*}[t!]
    \centering    
    \includegraphics[width = \linewidth]{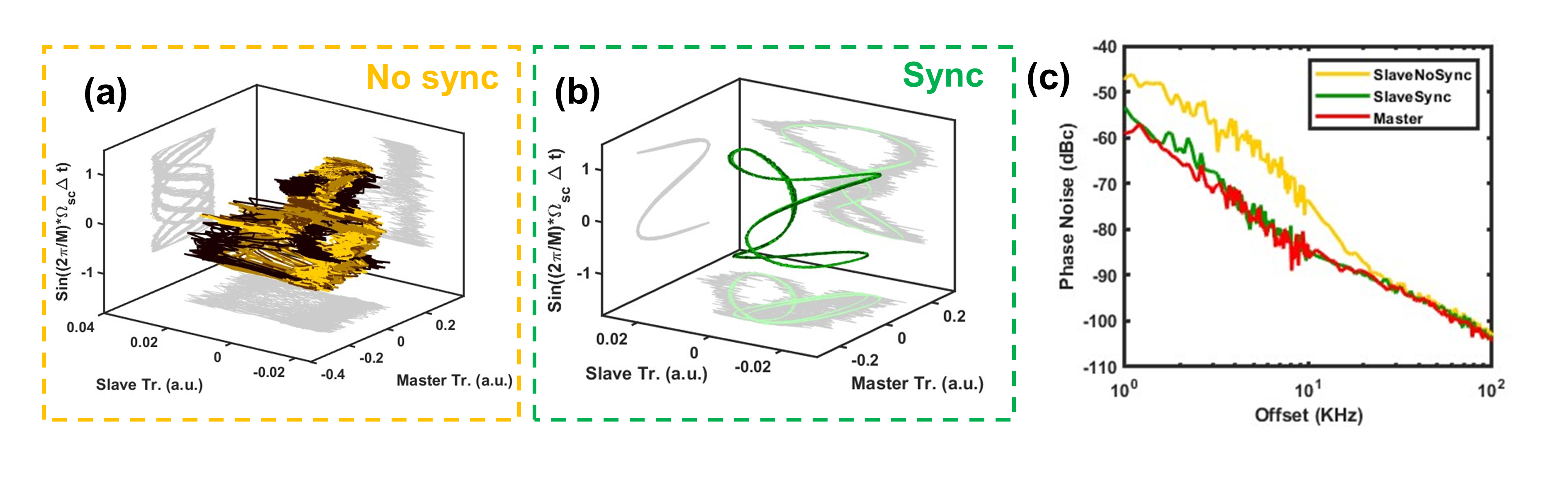}
    \caption{Temporal dynamics and phase noise of OM oscillators in the synchronization and free running regime. \textbf{(a)} Poincaré map of the recorded temporal traces of the free running slave using an stroboscopic technique with a sampling frequency of $f_{sync}/3$. Each colored curve corresponds to a different value of the initial delay ($\Delta t$). Projections in the different 2-dimensional planes are shown in grey. \textbf{(b)} Same representation for the case when the feedback amplification is above the synchronization threshold (20 dB). A fit (green) is performed to the raw data of the slave oscillation (grey).  \textbf{(c)} Phase noise of the master oscillator (red) and the free running and synchronized slave oscillator  (yellow and green, respectively) as a function of the frequency offset.}
    \label{fig: temporal}
\end{figure*}
Synchronization is observed above a certain amplification threshold value of 15 dB, given that the mechanical frequency of the slave is locked to that of the master one and most of the sidebands disappear. Two wide side-bands remain at a beating frequency of ($\Omega_M$ - $\Omega_S$)/2$\pi$, which are clear signatures of master-slave synchronization that have been reported in previous works addressing synchronization of photonic cavities \cite{Bagheri}. Their origin lies in the thermal force noise acting on the slave, which tends to push away the slave dynamics from the synchronization limit cycle. These experimental results have been compared with numerical simulations performed using a model based in the SP equations coupled to harmonic mechanical oscillators, showing a good qualitative agreement (see Supplementary Section S3). The measured synchronization mechanism resembles the suppression of the natural dynamics route described by Balanov \cite{Balanov} for a Van der Pol self-sustained oscillator under the actuation of a harmonic external force. Under this mechanism, the synchronization region is entered at relatively large amplitudes of forcing in comparison to a phase-locking mechanism, being one of the main characteristics of its route the absence of frequency pulling. Indeed, synchronization by suppression appears if the separation between the natural frequencies of the oscillators is rather large, which is indeed our case ($\Delta f$ = 0.7 MHz, i.e., about 0.7\%). As it is shown in Fig. \ref{fig: res1}b, the M = 3 mechanical lasing slave dynamics is preserved even after the threshold for synchronization. Thus, as mentioned earlier, this is an evidence that the slave is just adapting its dynamics to synchronize with the modulation generated by the master and hence discards resonant forcing. 
The intensity of the RF peaks appearing in the slave signal at the natural frequencies of each OMO has been analysed in Fig. \ref{fig: res1}d as a function of the feedback stage amplification. As expected, the RF peak associated with the slave remains constant until it sharply declines after synchronization. In the case of the master RF peak there is a linear relation with feedback amplification with a slope near to one, which indicates that the response of the modulation amplitude to the amplification/attenuation stage is linear. Interestingly, this linear tendency is altered with an abrupt increase of the master RF peak signal when the transition to synchronization occurs. This effect is linked to the transfer of self-sustained oscillation energy from the slave frequency to the master frequency, further confirming synchronization rather than forced oscillation. It is also worth mentioning that, as shown in Fig. \ref{fig: res1}c, even in the absence of external feedback there is a weak RF peak associated to the dynamics of the master, which is associated to a subtle mechanical cross-talk between the OMOs through the frame surrounding them. 
This interaction, in addition to being negligible compared to the amplitude of the lasing mechanical motion, is not in phase with the one introduced in the slave OMO using the external feedback mechanism. Therefore its contribution to the synchronization mechanism can be neglected.
\\
The temporal behaviour of the transmitted signals of both OMOs have been analyzed in the oscilloscope by recording traces of 800 ns using the slave OMO signal as the trigger. Fig. \ref{fig: temporal} shows the data represented as a Poincaré map, where the z-axis has been chosen to be $\sin \left(\frac{2\pi\Omega_{sync}}{3} \Delta t \right)$ to illustrate the trajectory in a three dimensional space. Below the synchronization threshold (Fig. \ref{fig: temporal}a) most of the phase space is filled by the traces, which is a clear indication that slave and master signals are not in sync. On the other hand, when synchronization occurs, the trajectory follows a closed curve in the phase diagram (Fig. \ref{fig: temporal}b). A fit was performed to the  oscillation trace of the slave to clearly observe the trajectory of the cycle. 
Finally, in Fig. \ref{fig: temporal}c the phase noise of the free running and synchronized slave OMO is reported and compared with that of the master. The synchronized slave reduces its phase noise at low frequencies until it becomes similar to that of the free running master OMO, which exhibit a value of -85 dBc/Hz at 10 KHz.  
\\
In conclusion, it has been unambiguously demonstrated a master-slave type of synchronization between two independent 1D optomechanical cavity self-sustained oscillators by introducing an external optical feedback mechanism. The route towards synchronization has been shown to be by suppression of the natural dynamics instead of the more standard phase-locking.  Furthermore, one dimensional cavities offer the advantage of being easy to integrate and less bulky compared to previous systems described in the literature. Even though in this work the oscillators are integrated in the same platform, the synchronization achieved by this method does not depend on the distance between them. In that way, the system presented here could be upscaled to networks of optomechanical oscillators interacting remotely. 

This work was supported by the MICINN projects ALLEGRO (Grants No. PID2021-124618NB-C22 and  PID2021-124618NB-C21) and MOCCASIN-2D (Grant No. TED2021-132040B-C21). A. M. acknowledges funding from the Generalitat Valenciana under grants IDIFEDER/2020/041 and IDIFEDER/2021/061.


%

\end{document}